\documentclass[11pt]{article}
\usepackage[dvips]{graphicx}
\usepackage{color}
\usepackage{amsfonts,amssymb,theorem,mathrsfs,times}

\usepackage{authblk}
\definecolor{wred}{rgb}{0,0.618,0.0}
\definecolor{wblue}{rgb}{.0,0.0,0.618}
\definecolor{wgreen}{rgb}{.0,0.618,0.0}

\usepackage[
      colorlinks=true,
      filecolor=black,
      linkcolor=wblue,
      citecolor=wgreen,
      pdfstartview=FitH,
        pdfpagemode=None,
        bookmarksopen=true,
      urlcolor=blue
      ]{hyperref}

\title{On  dRGT massive gravity with degenerate reference metrics}
\author[a, b]{Li-Ming Cao\footnote{Email: caolm@ustc.edu.cn}}
\affil[a]{Interdisciplinary Center for Theoretical Study,\ \

University of Science and Technology of China, Hefei, Anhui 230026, China}
\affil[b]{State Key Laboratory of Theoretical Physics,

Institute of Theoretical Physics,
 Chinese Academy of Sciences, Beijing 100190, China}

\affil[c]{  Department of Physics, 
National Taiwan University, Taipei 10617, Taiwan}

\author[a]{Yuxuan Peng\footnote{Email:  yxpeng@mail.ustc.edu.cn}}

\author[c]{Yun-Long Zhang\footnote{Email: zhangyunlong@ntu.edu.tw}}

\date{}                     
\newcommand{\bm}[1]{\mbox{\boldmath{$#1$}}}

\begin{document}

\maketitle
\begin{abstract}
In dRGT massive gravity, to get the equations of motion, the square root tensor is assumed to be invertible in the variation of the action. However, this condition can not be fulfilled when the reference metric is  degenerate.  This implies that the resulting equations of motion might be different from the case where  the reference metric has full rank. In this paper, by generalizing the Moore-Penrose inverse to the cases of symmetric tensors on  Lorentz manifolds, we get the equations of motion of the theory with a degenerate reference metric. It is found that the  equations of motion are a little bit different from those in the  non-degenerate cases. Based on the result of the equations of motion, for the $(2+n)$-dimensional solutions with the symmetry of $n$-dimensional maximally symmetric space, we  prove a generalized Birkhoff  theorem in the case where the degenerate reference metric has rank $n$, i.e., we  show that the solutions must be Schwarzschild-type or Nariai-Bertotti-Robinson-type under the given assumptions.

\end{abstract}

\newpage
{\tableofcontents}

\section{Introduction}

The concept massive gravity can be dated back to the early work \cite{Fierz:1939ix} introducing a mass term to the graviton of the linearized Einstein gravity by Fierz and Pauli. However, as a linear theory it suffered from the so-called vDVZ discontinuity\cite{vanDam:1970vg,Zakharov:1970cc}, i.e. the failure to resemble Einstein gravity while the graviton mass is approaching zero. This problem was solved by the Vainshtein mechanism\cite{Vainshtein:1972sx}. Meanwhile, most non-linear extensions of the Fierz-Pauli massive gravity had the problem of the Boulware--Deser (BD) ghost~\cite{Boulware:1973my}. There are many ways to avoid the ghost such as the DGP model\cite{Dvali:2000hr,Dvali:2000rv,Dvali:2000xg} and the "new massive gravity"\cite{Bergshoeff:2009hq}, and recently, the dRGT theory\cite{deRham:2010ik,deRham:2010kj} by de Rham, Gabadadze, and Tolley.
The dRGT theory was proven to be ghost-free in the ADM formalism by \cite{Hassan:2011hr,Hassan:2011tf}. For more details the readers can refer to the reviews \cite{Hinterbichler:2011tt, deRham:2014zqa}. We just mention that the dRGT massive gravity theory includes a non-dynamical ``reference metric" $f_{ab}$ in its formulation, thus breaking the diffeomorphism invariance which is preserved by Einstein gravity. 

Recently the nonlinear massivie gravity had been studied extensively in the context of AdS/CFT correspondence on the realization of momentum dissipation\cite{Vegh:2013sk,Davison:2013jba,Blake:2013bqa} in the boundary field theory due to the absence of the translational symmetry on the boundary dual to the diffeomorphism invariance in the bulk. The paper \cite{Cao:2015cza} constructed the local counterterms of this theory to study the renormalized thermodynamic quantities and the renormalized boundary stress energy tensor. 
The thermodynamics and phase transition of the massive gravity model were studied in the paper \cite{Cai:2014znn} and the case in the extended phase space was studied in \cite{Xu:2015rfa}. With some special assumptions, the authors of the paper \cite{Hu:2015xva} derived the Misner-Sharp mass and they also  studied the thermodynamics.
The dRGT theory is also studied in depth in cosmology. Homogeneous and isotropic solutions and other issues such as self-acceleration are reviewed in the paper \cite{DeFelice:2013bxa} and the reader can refer to this paper for more details.

Interestingly, there is an alternative form of the massive gravity model, applying the so-called St\"{u}ckelberg fields\cite{deRham:2011rn} appearing as scalar fields instead of the reference metric $f_{ab}$. Some researchers studied the holographic conductivity by regarding them as dynamical fields\cite{Alberte:2014bua}. 
In this St\"{u}ckelberg language the massive gravity theory appears to be diffeomorphism invariant. Similar models (not dRGT theory) with more general scalar field terms are thoroughly studied by \cite{Baggioli:2014roa,Alberte:2015isw}.

In this paper we focus on a more fundamental issue of the massive gravity theory. This is the derivation of the equations of motion.
In this theory the reference metric $f_{ab}$ is somewhat arbitrary. When the matrix of this tensor in a certain basis is  non-degenerate it is not hard to derive the equations of motion by using the inverse of the square root tensor ${\bm \gamma}_a{}^b$ which is defined by ${\bm \gamma}_a{}^c {\bm \gamma}_c{}^b=f_a{}^b$. 
The paper \cite{Hassan:2011vm} offers a good explanation for this.
However, many papers, for example, \cite{Vegh:2013sk, Cai:2014znn} and the paper \cite{Zhang:2015nwy} which studies the ghost problem, deals with degenerate reference metrics.
When $f_{ab}$ is degenerate so is the tensor ${\bm \gamma}_a{}^b$ and one should be careful with the concept ``inverse". In this paper we apply the Moore-Penrose pseudoinverse \cite{Moore:1920, Bjerhammar:1951, Penrose:1955} of the tensor ${\bm \gamma}_a{}^b$ and derive the equations of motion for degenerate ${\bm \gamma}_a{}^b$s. 
We show that if the Moore-Penrose pseudoinverse exists then the equations of motion appear in the same form as those for the nondegenerating ${\bm \gamma}_a{}^b$s. Moreover, if ${\bm \gamma}_a{}^b$ has no such a pseudoinverse one can perform a ``translation" and get an equivalent tensor which has one. This is due to an uncertainty of the matrix square root. Then the equations of motion can be derived for any degenerate ${\bm \gamma}_a{}^b$. With the equations of motion at hand we prove a Birkhoff type theorem of massive gravity. By choosing a rank-$n$ reference metric, for the $(2+n)$-dimensional solutions  with the symmetry of $n$-dimensional maximally symmetric space, we show that the solutions are either Schwarzschild-type or Nariai-Bertotti-Robinson-type.

In usual matrix theory in linear algebra, any matrix has an unique Moore-Penrose inverse,  so one might  wonder why these authors have arrived at  the place where the pseudoinverses do not exist? This is because we are considering
matrix in indefinite linear algebra with Minkowski inner product. 
For two vector spaces $V$ and $W$ with Euclidean inner products,  any  matrix  $A: V\rightarrow W$ has a
Moore-Penrose inverse $A^+:W\rightarrow V$ which can be defined by the following equations
\begin{eqnarray}
AA^+A=A\, , \qquad  A^+AA^+=A^+\, ,
\end{eqnarray}
\begin{eqnarray}
\label{transposecondition}
(AA^+)^T=AA^+\, ,\quad (A^+A)^T=A^+A\, ,
\end{eqnarray}
where $T$ denotes the transpose of the matrix (if complex numbers appear, the transpose should be understood as conjugate transpose).
In this definition, we have to be careful about the transpose of the matrix.
Here, we give some discussions on the meaning of the transpose of the matrix.
Generally, the transpose of a linear mapping $f: V\rightarrow W$ is a mapping $f^t:W^*\rightarrow V^*$ such that for each $w\in W^*$ and each $v\in V$, we have
\begin{equation}
(f^t w) (v) =  w (f v)\, ,
\end{equation}
or in indices form
\begin{equation}
\label{transposeindices}
(f^t)_a{}^{i}w_iv^a =w_i f^i{}_av^a\, .
\end{equation}
In the above equations, only the notion of the dual of a linear space has been used. This kind of definition of transpose is popularly used in differential geometry. With this definition in hand, let us investigate
the transpose conditions in equations  (\ref{transposecondition}).  Notice  that $(A^+A)$ is a mapping from $V$ to $V$, so the transpose of $(A^+A)$ is a mapping from $V^*$ to $V^*$. Without some relation between $V$ and $V^*$, we can not
identify these two mappings.  This simple investigation implies that some structure has been used in the condition  (\ref{transposecondition}). This structure is   nothing  but the Euclidean inner product or Euclidean metric of
the space $V$. Assume that $g$ is the Euclidean metric on $V$, then we have an identification between $V$ and $V^*$, i.e.,  for any $v^a\in V$, $v_a=g_{ab}v^b$ is a dual vector in $V^*$.
 Similarly, to understand that the transpose of $(AA^+)$ is equal to $(AA^+)$, we have to introduce an Euclidean metric, denoted by $h$, on $W$.
After introducing the metrics, the indices can be raised and  lowered freely with the order unchanged. For example, in Eq.(\ref{transposeindices}), the indices $a$ and
$i$ can be raised and lowered by metric $g$ and $h$ respectively. However, the order of $a$ and $i$ in $f$ or $f^t$  can not be changed in general.  Based on this discussion, it is not hard to find that the transpose $T$ in condition (\ref{transposecondition})
should be understood as
\begin{equation}
(A^+A)^T = g^{-1}(A^+A)^t  g = A^+ A\, ,
\end{equation}
and
\begin{equation}
(AA^+)^T = h^{-1} (AA^+)^t h = A A^+\, .
\end{equation}
Of course, in usual matrix theory, the Euclidean metrics can be viewed as  unit matrices, and the transpose $t$ is the same as $T$ in some sense.  However, if we are considering the space with
Lorentzian signature, $g$ or $h$ can not be omitted any more.
Now, let $V$ and $W$  to be  two spaces with Minkowski metrics $g$ and $h$ respectively, then by adding all possible indices, the generalized Moore-Penrose inverse
is defined as
\begin{eqnarray}
\label{generalizedMPI1}
A^i{}_c(A^+)^c{}_jA^j{}_a=A^i{}_a\, , \qquad (A^+)^a{}_jA^j{}_c(A^+)^c{}_i=(A^+)^a{}_i\, ,
\end{eqnarray}
and
\begin{eqnarray}
\label{generalizedMPI2}
A_{j}{}^a(A^+)_a{}^{i}=A^i{}_a(A^+)^a{}_j\, ,\qquad  (A^+)_b{}^iA_i{}^a=(A^+)^a{}_iA^i{}_b\, .
\end{eqnarray}
This formalism is quite simple and can be easily applied in gravity theory. 
If $V$ and $W$ are both defined on complex  fields, the complex conjugate should be added to the left sides of the equations (\ref{generalizedMPI2}). Since the metric has Lorentzian signature, this generalized Moore-Penrose   inverse does not always exist. This point will be found in Sec.{\ref{degenerate deri}}.
Actually, the generalized Moore-Penrose inverse here   is closely related to
the so-called weighted Moore-Penrose inverse in the community of mathematics.
For example,  the weighted Moore-Penrose inverse in general indefinite linear algebra has been studied in reference~\cite{SG}.

This paper is organized as follows. Firstly, we review the procedure of deriving the equations of motion for massive gravity theory with a   non-degenerate  reference metric in the next section. Then we give the derivation for a degenerate reference metric and discuss the uncertainty of the square root tensor in section \ref{degenerate deri}. Different types of the square root tensors and the physical validity for each type are analyzed in section \ref{energy condi}. Based on the equations of motion we derive, in section \ref{Birk theo} we show a Birkhoff-type theorem  in the case where the degenerate reference metric has rank $n$ and the spacetime has a maximal symmetric subspace of dimension $n$. In addition, the form of the square root tensor is restricted as shown in the proposition in section \ref{Prop}. In the Appendix we give the proof of this proposition.

\section{  Equations of motion with non-degenerate reference metrics}
The action of the     dRGT  massive gravity theory  in   $d$ dimensions  is given by
\begin{equation}\label{actionmassive}
S=\frac{1}{2\kappa^2}\int \mathrm{d}^dx \sqrt{-g}\Big[R[g]+m^2 \sum_{i=0}^d c_i\mathcal{U}_i(g,f)+\mathrm{matter~part}\Big]\, ,
\end{equation}
where $R[g] $ is the scalar curvature of the spacetime $(M, g_{ab})$.  The symmetric tensor  $f_{ab}$ is the so-called ``reference metric" of the theory, and $c_0\, ,c_1\, ,\cdots\, ,c_d$ are constants.
The scalars  $\mathcal{U}_i$s  are defined by a recursive relation starting from $\mathcal{U}_0=1$, i.e.,
\begin{equation}
\mathcal{U}_i=-(i-1)! \sum_{j=1}^{i}\frac{(-1)^j}{(i-j)!} \mathrm{Tr}({\bm \gamma}^j) \mathcal{U}_{i-j}\, ,\qquad i\ge 1\, .
\end{equation}
The term including $\mathcal{U}_0$ can be viewed as the cosmological constant term of the theory, and $\mathcal{U}_i$s with $i\ge 1$ have   the following forms
\begin{equation}
\mathcal{U}_1=\mathrm{Tr}{\bm \gamma}\, ,
\end{equation}
\begin{equation}
\mathcal{U}_2= (\mathrm{Tr}{\bm \gamma})^2-\mathrm{Tr}({\bm \gamma}^2)\, ,
\end{equation}
\begin{equation}
\mathcal{U}_3=(\mathrm{Tr}{\bm \gamma})^3-3(\mathrm{Tr}{\bm \gamma})\mathrm{Tr}({\bm \gamma}^2)+2 \mathrm{Tr}({\bm \gamma}^3)\, ,\end{equation}
\begin{equation}
\mathcal{U}_4=(\mathrm{Tr}{\bm \gamma})^4-6(\mathrm{Tr}{\bm \gamma})^2\mathrm{Tr}({\bm \gamma}^2)+8 (\mathrm{Tr}{\bm \gamma})\mathrm{Tr}({\bm \gamma}^3)+3\mathrm{Tr}({\bm \gamma}^2)\mathrm{Tr}({\bm \gamma}^2)-6\mathrm{Tr}({\bm \gamma}^4)\, .
\end{equation}
The symbol $\mathrm{Tr}\mathcal{A}$ denotes the trace of some tensor $\mathcal{A}$. The square root tensor ${\bm \gamma}_{ab}$ is usually assumed to be
a symmetric tensor on the spacetime $(M, g_{ab})$, and satisfies
\begin{equation}
{\bm \gamma}_a{}^c{\bm \gamma}_c{}^b= f_{ac}g^{cb}\, ,
\end{equation}
and ${\bm \gamma}^i$ is defined by the contraction of $i$ copies of  ${\bm \gamma}$, i.e.,
\begin{equation}
({\bm \gamma}^i)_{a}{}^b= {\bm \gamma}_{a}{}^{c_1}{\bm \gamma}_{c_1}{}^{c_2}\cdots {\bm \gamma}_{c_i}{}^b\, ,
\end{equation}
Usually, we set $$({\bm \gamma}^0)_{a}{}^b=\delta_a{}^b\, ,\quad ({\bm \gamma}^0)_{ab}=g_{ab}\, .$$ The trace of the square root tensor can be viewed as the contraction of the metric and ${\bm \gamma}$, i.e., we have $$\mathrm{Tr}({\bm \gamma}^i)=g^{ab}({\bm \gamma}^i)_{ab}\, .$$

In the variation of the action, the most important part is the variation of the trace of the square  root tensor ${\bm \gamma}_{ab}$, i.e., the variation of $\mathrm{Tr}{{\bm \gamma}}$. To make the question more transparent, the variation of this term
can be done as follows
\begin{eqnarray}
2\delta \mathrm{Tr}{\bm \gamma}&=& \delta_a{}^b\delta {\bm \gamma}_b{}^a+\delta_a{}^b\delta{\bm \gamma}_b{}^a\nonumber\\
&=&({\bm \gamma}^{-1})_a{}^c{\bm \gamma}_c{}^b\delta {\bm \gamma}_b{}^a+ {\bm \gamma}_a{}^c({\bm \gamma}^{-1})_c{}^b\delta{\bm \gamma}_b{}^a\nonumber\\
&=&({\bm \gamma}^{-1})_a{}^c{\bm \gamma}_c{}^b\delta {\bm \gamma}_b{}^a+({\bm \gamma}^{-1})_a{}^c \delta{\bm \gamma}_c{}^b{\bm \gamma}_b{}^a\nonumber\\
&=&({\bm \gamma}^{-1})_a{}^c\delta ({\bm \gamma}_c{}^b{\bm \gamma}_b{}^a)\nonumber\\
&=&({\bm \gamma}^{-1})_a{}^c\delta (f_{cb}g^{ba})\nonumber\\
&=&({\bm \gamma}^{-1})_a{}^cf_{cb}\delta g^{ba}\nonumber\\
&=&({\bm \gamma}^{-1})_a{}^c{\bm \gamma}_c{}^d{\bm \gamma}_{db}\delta g^{ba}\nonumber\\
&=&{\bm \gamma}_{ab}\delta g^{ab}\, ,
\end{eqnarray}
where the reference metric $f_{ab}$ is fixed, i.e., $\delta f_{ab}=0$. The variation of $\mathrm{Tr}{\bm \gamma}^i\, ,i\ge 2$ is quite simple, and  then the variation of $\mathrm{Tr}{\bm \gamma}^i$ has a form
\begin{equation}
 \delta \mathrm{Tr}{\bm \gamma}^{i}=\frac{i}{2}({\bm \gamma}^{i})_{ab}\delta g^{ab}\, , \quad i=1\, ,\cdots d\, .
\end{equation}
Based on the above variation, one gets the equations of motion for the massive gravity
\begin{equation}
\label{eom1}
G_{ab} + m^2 X_{ab}=0\, ,
\end{equation}
where
\begin{equation}
2 X_{ab}= -\sum_{i=0}^d  c_i \Big[ \sum_{j=0}^{i}(-1)^{i-j}(i!)(j!)^{-1}\mathcal{U}_j({\bm \gamma}^{i-j})_{ab}\Big]\, .
\end{equation}
For example, when $d=4$, we have
\begin{eqnarray}
\label{Xab}
2X_{ab}&=&- c_0 g_{ab} -c_1(\mathcal{U}_1 g_{ab}-{\bm \gamma}_{ab}) - c_2 (\mathcal{U}_2 g_{ab}-2\mathcal{U}_1{\bm \gamma}_{ab}+2 ({\bm \gamma}^2)_{ab})\nonumber\\
&&-c_3\big(\mathcal{U}_3g_{ab}-3\mathcal{U}_2 {\bm \gamma}_{ab}+6\mathcal{U}_1({\bm \gamma}^2)_{ab}-6({\bm \gamma}^3)_{ab}\big)\nonumber\\
&&-c_4\big(\mathcal{U}_4g_{ab}-4\mathcal{U}_3 {\bm \gamma}_{ab}+12\mathcal{U}_2({\bm \gamma}^2)_{ab}-24 \mathcal{U}_1({\bm \gamma}^3)_{ab}+24({\bm \gamma}^4)_{ab}\big)\, .\nonumber\\
\end{eqnarray}
Obviously,  the inverse of the square root tensor, i.e., ${\bm \gamma}^{-1}$, is important in this procedure. If the square root tensor ${\bm \gamma}_{ab}$ is degenerate, we do not know whether these equations are available or not.

\section{ Equations of motion with degenerate reference metric}\label{degenerate deri}


In this section,  we firstly generalize the pseudoinverse of degenerate  square root tensor ${\bm \gamma}_{ab}$, and then derive the equations of motion with both of the pseudoinvertible and non-pseudoinvertible case. 


\subsection{Generalized pseudoinverse}
When  the reference metric  $f_{ab}$ is degenerate, the square root tensor ${\bm \gamma}_{ab}$ is   also degenerate.
Here, we give a deduction of the equations of motion. Similar to the non-degenerate cases, we have to consider the variation of the trace of the tensor ${\bm \gamma}_{ab}$. However,
the situation is quite non-trivial because now the square root tensor is degenerate in general, and the procedure in the previous section is not available.

Usually, the square root tensor ${\bm \gamma}_a{}^b$ can be viewed as a matrix, or  a linear mapping from $T_pM$ to $T_pM$ ($p\in M$), so  the Moore-Penrose pseudoinverse\cite{Moore:1920, Bjerhammar:1951, Penrose:1955} of ${\bm \gamma}_a{}^b$
might be useful to get the variation of $\mathrm{Tr}{\bm \gamma}$.  Probably, we can get this variation by applying the  Moore-Penrose inverse ${\bm \gamma}^{+}$ instead of the usual  inverse  ${\bm \gamma}^{-1}$.
However, since  $T_pM$ is a linear space with Minkowski inner product, the  Moore-Penrose inverse in Euclidean space is not enough, we have to generalized the pseudoinverse to the linear space with
Minkowski inner product.

Since the square root tensor is symmetric, it is enough to consider the Moore-Penrose pseudoinverse of symmetric tensors. Generally, the symmetric tensor on $(2+n)$-dimensional spacetime can be classified into
four Segre types~\cite{exactsolutions, Santos:1995kt}, i.e., $[1,1\cdots1]$, $[21\cdots 1]$, $[31\cdots 1]$, and $[z\bar{z}1\cdots 1]$. These four types can be put into more familiar form by using principal eigenvectors of the tensor. This means
that ${\bm \gamma}_{ab}$ can be expressed in canonical forms by introducing appropriate null frames. As usual,  we can introduce null frames like
$$\{\ell_a\, ,n_a\, ,x^{(i)}_a\, ,\quad i=1\, ,\cdots\, ,n\}\, ,$$
where
\begin{eqnarray}
&&\ell_a\ell^a=n_an^a=0\, ,\qquad \ell_an^a=-1\, ,\nonumber\\
&&\ell^ax^{(i)}_a=n^ax^{(i)}_a=0\, ,\quad g^{ab} x_a^{(i)}x_{b}^{(j)}=\delta^{ij}\, .
\end{eqnarray}
Actually, sometimes it is also convenient to define orthogonal frames by
$$\{t_a\, ,z_a\, ,x^{(i)}_a\, ,\quad i=1\, ,\cdots\, ,n\}\, ,$$
where
\begin{equation}
t_a=\frac{1}{\sqrt{2}}(\ell_a + n_a)\, ,\quad z_a=\frac{1}{\sqrt{2}}(\ell_a-n_a)\, .
\end{equation}
Obviously,  $t^a$ is a timelike unit vector, and $z^a$ is a spacelike unit vector.  Here, the null vectors $\ell^a$ and $n^a$ are both future-pointing, i.e., we have $\ell_at^a<0$ and $n_at^a<0$.
 Of course, these frames  (null or orthogonal) are not unique, Lorentz transformations can be applied to
these frames to get other frames which also satisfy the above conditions. Actually, by using these Lorentz transformations, according to the Sygre types, symmetric tensors can be put into simple forms as follows
\begin{itemize}
\item [$(1).$]Type $[1,1\cdots 1]$. In this type, ${\bm \gamma}_a{}^b$ has $(2+n)$ real eigenvalues and $(2+n)$ eigenvectors. By choosing an appropriate frame, the symmetric tensor can be expressed as
\begin{equation}
\qquad {\bm \gamma}_{ab}=- 2\alpha_1 \ell_{(a}n_{b)} + \alpha_2 (\ell_a\ell_b + n_an_b) + \sum_{i=1}^{n}\alpha_{i+2}~ x^{(i)}_ax^{(i)}_b\, .
\end{equation}
In the orthogonal frame, a tensor of this type has a diagonal form,
\begin{equation}
{\bm \gamma}_{ab}=-\lambda_1t_at_b + \lambda_2 z_az_b + \sum_{i=1}^{n}\lambda_{i+2}~ x^{(i)}_ax^{(i)}_b\, ,
\end{equation}
where $\lambda_1=\alpha_1-\alpha_2$, $\lambda_2=\alpha_1+\alpha_2$, and $\lambda_{i+2}=\alpha_{i+2}\, ,i=1\, ,\cdots\, ,n$. Obviously, $\lambda_i\, ,i=1\, ,\cdots\, ,n+2$ are the eigenvalues of ${\bm \gamma}_a{}^b$, and
$t_a$, $z_a$, and $x_a^{(i)}$ are the corresponding eigenvectors.

\item [$(2).$]Type $[21\cdots 1]$ has $(1+n)$ eigenvalues and $(1+n)$ eigenvectors and can be expressed as
\begin{equation}
{\bm \gamma}_{ab}= -2\beta_1 \ell_{(a}n_{b)} +\beta_2\ell_a\ell_b + \sum_{i=1}^{n}\beta_{i+2}~ x^{(i)}_ax^{(i)}_b\, ,
\end{equation}
where $\beta_2=\pm 1$.  In this null frame, the eigenvectors are $\ell_a$ and $x_a^{(i)}$, and associated eigenvalues are $\beta_1$ and $\beta_{i+2}$. Here, $i=1\, ,\cdots\, ,n$.

\item [$(3).$]Type $[31\cdots 1]$ has $n$ eigenvectors, and has a form
\begin{equation}
{\bm \gamma}_{ab}=- 2\gamma_1 \ell_{(a}n_{b)} +2 \gamma_2 \ell_{(a}x^{(1)}_{b)} + \sum_{i=2}^{n}\gamma_{i+1}~ x^{(i)}_ax^{(i)}_b\, ,
\end{equation}
where $\gamma_2=1$. One of the eigenvector is the null vector $\ell_a$, and others are spacelike, i.e., $x_a^{(i)}\, ,i=2\, ,\cdots, n$.

\item [$(4).$]Type $[z\bar{z}\cdots 1]$ has two complex eigenvalues and $n$ real eigenvalues.  In an appropriate null frame, it has a form
\begin{equation}
{\bm \gamma}_{ab}=- 2\sigma_1 \ell_{(a}n_{b)} + \sigma_2 (\ell_a\ell_b - n_an_b) + \sum_{i=1}^{n}\sigma_{i+2}~ x^{(i)}_ax^{(i)}_b\,,
\end{equation}
where $\sigma_2\ne 0$. The  eigenvectors are $\ell_a \pm i n_a$ with complex eigenvalues $-(\sigma_1\pm i \sigma_2)$. Other eigenvectors are real spacelike vectors. In the complex orthogonal frame, this type can be expressed as
\begin{equation}
{\bm \gamma}_{ab}=-i\kappa_1 t_at_b + i \kappa_2 z_az_b + \sum_{i=1}^{n}\kappa_{i+2}~ x^{(i)}_ax^{(i)}_b\, ,
\end{equation}
where $t^a$ and $z^a$ are two complex vectors
\begin{equation}
t_a=\frac{1}{\sqrt{2}}(\ell_a + i n_a)\, ,\quad z_a= \frac{1}{\sqrt{2}}(\ell_a - i n_a)\, ,
\end{equation}
and $\kappa_1=-\sigma_1+i\sigma_2$, $\kappa_2=-\sigma_1-i\sigma_2$, and $\kappa_i=\sigma_{i+2}\, ,i=1\,,\cdots\,, n$.
\end{itemize}
In the above equations, $\alpha_i$, $\beta_i$, $\gamma_i$, and $\sigma_i$ are real.

According to the discussion in the introduction,  Eqs.(\ref{generalizedMPI1}) and (\ref{generalizedMPI2}) can be used to define the pseudoinverse of the symmetric tensors.  The pseudoinverse ${\bm \gamma}^+$ of the square root tensor is defined by the conditions
\begin{eqnarray}
&&{\bm \gamma}_a{}^c({\bm \gamma}^+)_c{}^d{\bm \gamma}_d{}^b={\bm \gamma}_a{}^b\, ,\nonumber\\
&&({\bm \gamma}^+)_a{}^c{\bm \gamma}_c{}^d({\bm \gamma}^+)_d{}^b=({\bm \gamma}^+)_a{}^b\, ,\nonumber\\
&&({\bm \gamma}^+)^b{}_c{\bm \gamma}^c{}_a=({\bm \gamma}^+)_a{}^c{\bm \gamma}_c{}^b\, ,\nonumber\\
&&{\bm \gamma}^b{}_c({\bm \gamma}^+)^c{}_a={\bm \gamma}_a{}^c({\bm \gamma}^+)_c{}^b\, .
\end{eqnarray}
The complex conjugate should be added in the left sides of the last two equations if the tensor has complex eigenvalues.
By this definition, we find the generalized Moore-Penrose inverses always exist for the type $[1,1\cdots 1]$ and type $[z\bar{z}1\cdots 1]$, and they have the forms
\begin{equation}
({\bm \gamma}^+)_{ab}=-\lambda_1^+~ t_at_b + \lambda_2^+~ z_az_b + \sum_{i=1}^{n} \lambda_{i+2}^+ ~ x^{(i)}_ax^{(i)}_b\, ,
\end{equation}
and
\begin{equation}
({\bm \gamma}^+)_{ab}=-i\kappa_1^+~ t_at_b +i\kappa_2^+~ z_az_b + \sum_{i=1}^{n} \kappa_{i+2}^+ ~ x^{(i)}_ax^{(i)}_b\, ,
\end{equation}
where for an arbitrary  number $\lambda$, we have defined
\[\lambda^+=
\left\{
\begin{array}{ll}
\lambda^{-1}& \quad \lambda\ne 0,\\
0&\quad \lambda=0\, .
\end{array}\right.\]
However, for the type $[21\cdots 1]$ and $[31\cdots 1]$, the generalized Moore-Penrose inverses only exist when $\beta_1\ne 0$ and $\gamma_1\ne 0$, and we have
\begin{equation}
({\bm \gamma}^+)_{ab}=-\beta_1^+ 2\ell_{(a}n_{b)} - \beta_2^+ (\beta_1^+)^2 \ell_a\ell_b + \sum_{i=1}^{n}\beta_{i+2}^+~ x^{(i)}_ax^{(i)}_b\, ,
\end{equation}
and
\begin{equation}
({\bm \gamma}^+)_{ab}= (\gamma_1^+)^3 \ell_a\ell_b - 2 \gamma_1^+ \ell_{(a}n_{b)}- \gamma_2^+(\gamma_1^+)^2 \ell_{(a}x^{(1)}_{b)} + \sum_{i=2}^{n}\gamma_{i+1}^+~ x^{(i)}_ax^{(i)}_b\, .
\end{equation}
These pseudoinverses are unique and also symmetric.
So the generalized Moore-Penrose inverses do not always exist. This is very different from the Euclidean case where the Moore-Penrose inverse of a matrix always exists and is unique.

It is not hard to find the common feature of the cases with vanishing $\beta_1$ and $\gamma_1$---the rank (as matrix) of the square root tensor ${\bm \gamma}_a{}^b$ is different from the rank of ${\bm \gamma}_a{}^c{\bm \gamma}_c{}^b$, i.e.,
\begin{equation}
\mathrm{rank}({\bm \gamma})\ne \mathrm{rank}({\bm \gamma}^2)\, .
\end{equation}
This can be found as follows: Since the rank of a linear mapping is  the dimension of its range, the investigation of the range of ${\bm \gamma}$ and ${\bm \gamma}^2$ is sufficient to get their ranks. Assuming $$v_a=a\ell_a+ bn_a+ e_i x_a^{(i)}\, ,$$
where $a\, ,b$, and $e_i$s are arbitrary numbers, we have
\begin{equation}
{\bm \gamma}_{a}{}^bv_b=(a\beta_1- 2 b \beta_2 )\ell_a + b \beta_1 n_a + \sum_{i=1}^n e_i \beta_{i+2} x_a^{(i)}\, ,
\end{equation}
and
\begin{equation}
{\bm \gamma}_a{}^c{\bm \gamma}_{c}{}^bv_b=(a\beta_1^2 - 2 b \beta_1\beta_2 )\ell_a + b \beta_1^2 n_a + \sum_{i=1}^n e_i \beta_{i+2}^2 x_a^{(i)}\, .
\end{equation}
Obviously, when $\beta_1\ne 0$,  the dimensions of  $\mathrm{range}({\bm \gamma})$ and $\mathrm{range}({\bm \gamma}^2)$ are equal. However, when $\beta_1=0$, the dimension of
$\mathrm{range}({\bm \gamma})$ is bigger than the dimension of $\mathrm{range}({\bm \gamma}^2)$, and we have
\begin{equation}
\mathrm{rank}({\bm \gamma})=1+\mathrm{rank}({\bm \gamma}^2)\, .
\end{equation}
This result is also correct for the case of type $[31\cdots 1]$ with vanishing $\gamma_1$. In the cases where generalized Moore-Penrose inverses exist, it is easy to find $\mathrm{rank}({\bm \gamma})=\mathrm{rank}({\bm \gamma}^2)$. This investigation suggests that the generalized pseudoinverses only exist  when the square root tensor and the reference metric have the same rank. Similar conclusion can also be found in~\cite{SG} with more general indefinite inner products.

\subsection{Pseudoinvertible square root tensor}

When the generalized Moore-Penrose inverse of ${\bm \gamma}$ exists, we can get the variation of $\mathrm{Tr}{\bm \gamma}$. Since we have
\begin{eqnarray}
&&\delta \mathrm{Tr}{\bm \gamma}=\delta \Big({\bm \gamma}_a{}^c({\bm \gamma}^+)_c{}^d{\bm \gamma}_d{}^a\Big)\nonumber\\
&&=({\bm \gamma}^+)_c{}^d{\bm \gamma}_d{}^a\delta{\bm \gamma}_a{}^c+ ({\bm \gamma}^+)_c{}^d\delta{\bm \gamma}_d{}^a{\bm \gamma}_a{}^c+{\bm \gamma}_a{}^c \delta({\bm \gamma}^+)_c{}^d{\bm \gamma}_d{}^a\nonumber\\
&&=({\bm \gamma}^+)_c{}^d\delta({\bm \gamma}_d{}^a{\bm \gamma}_a{}^c)-{\bm \gamma}_a{}^b ({\bm \gamma}^+)_b{}^c \delta{\bm \gamma}_c{}^d ({\bm \gamma}^+)_d{}^e{\bm \gamma}_e{}^a\nonumber\\
&&=({\bm \gamma}^+)_c{}^d\delta(f_{db}g^{bc})-{\bm \gamma}_a{}^b ({\bm \gamma}^+)_b{}^c \delta{\bm \gamma}_c{}^d ({\bm \gamma}^+)_d{}^e{\bm \gamma}_e{}^a\, ,
\nonumber\\
&&=({\bm \gamma}^+)_c{}^df_{db}\delta g^{bc}-{\bm \gamma}_a{}^b ({\bm \gamma}^+)_b{}^c \delta{\bm \gamma}_c{}^d ({\bm \gamma}^+)_d{}^e{\bm \gamma}_e{}^a\, ,
\nonumber\\
&&=({\bm \gamma}^+)_c{}^d{\bm \gamma}_d{}^e{\bm \gamma}_{eb}\delta g^{bc}-{\bm \gamma}_a{}^b ({\bm \gamma}^+)_b{}^c \delta{\bm \gamma}_c{}^d ({\bm \gamma}^+)_d{}^e{\bm \gamma}_e{}^a\, ,
\end{eqnarray}
where we have used $\delta f_{ab}=0$ and the relation
\begin{equation}
 {\bm \gamma}_a{}^c \delta({\bm \gamma}^+)_c{}^d{\bm \gamma}_d{}^a=-{\bm \gamma}_a{}^b ({\bm \gamma}^+)_b{}^c \delta{\bm \gamma}_c{}^d ({\bm \gamma}^+)_d{}^e{\bm \gamma}_e{}^a\, .
\end{equation}
Considering the symmetry of the square root tensor and its pseudo inverse (if exists), we have
\begin{equation}
{\bm \gamma}_a{}^b ({\bm \gamma}^+)_b{}^c ={\bm \gamma}^c{}_b ({\bm \gamma}^+)^b{}_a ={\bm \gamma}_b{}^c ({\bm \gamma}^+)_a{}^b = ({\bm \gamma}^+)_a{}^b{\bm \gamma}_b{}^c\, .
\end{equation}
Substituting this relation into the variation of $\mathrm{Tr}{\bm \gamma}$, we have
\begin{eqnarray}
&&2\delta \mathrm{Tr}{\bm \gamma}=2{\bm \gamma}_c{}^d({\bm \gamma}^+)_d{}^e{\bm \gamma}_{eb}\delta g^{bc}\nonumber\\
&&-({\bm \gamma}^+)_d{}^e{\bm \gamma}_e{}^a({\bm \gamma}^+)_a{}^b {\bm \gamma}_b{}^c \delta{\bm \gamma}_c{}^d -({\bm \gamma}^+)_e{}^a{\bm \gamma}_a{}^b ({\bm \gamma}^+)_b{}^c \delta{\bm \gamma}_c{}^d {\bm \gamma}_d{}^e\nonumber\\
&&
=2{\bm \gamma}_c{}^d({\bm \gamma}^+)_d{}^e{\bm \gamma}_{eb}\delta g^{bc}-({\bm \gamma}^+)_d{}^b {\bm \gamma}_b{}^c \delta{\bm \gamma}_c{}^d -({\bm \gamma}^+)_d{}^b \delta{\bm \gamma}_b{}^c {\bm \gamma}_c{}^d\nonumber\\
&&
=2{\bm \gamma}_c{}^d({\bm \gamma}^+)_d{}^e{\bm \gamma}_{eb}\delta g^{bc}-({\bm \gamma}^+)_d{}^b \delta({\bm \gamma}_b{}^c {\bm \gamma}_c{}^d)\nonumber\\
&&={\bm \gamma}_c{}^d({\bm \gamma}^+)_d{}^e{\bm \gamma}_{eb}\delta g^{bc}\nonumber\\
&&={\bm \gamma}_{ab}\delta g^{ab}\, .
\end{eqnarray}
So the result is the same as the case with a non-degenerate reference metric. By this result, we find that the equations of motion are the same as those in the non-degenerate case
if the generalized Moore-Penrose inverse of the square root tensor exists.

\subsection{Non-pseudoinvertible square root tensor}

In the cases where the generalized Moore-Penrose inverses do not exist, we have no general way to get the equations of motion. However, note that the action and the reference metric are both invariant
under the translation
\begin{equation}
{\bm \gamma}_{ab}\rightarrow {\bm \gamma}'_{ab}= {\bm \gamma}_{ab}-\Phi \ell_a\ell_b
\end{equation}
if the square root tensor is degenerate along the null direction $\ell^a$, i.e., ${\bm \gamma}_{ab}\ell^b=0$. Here $\Phi$ is an arbitrary function. So, in this case, the translation is a kind of symmetry of the theory, and the equations of motion should be invariant under this translation.

For the type $[21\cdots 1]$ with vanishing $\beta_1$, the square root tensor ${\bm \gamma}_{ab}$ is degenerate along the null direction $\ell^a$. After the above translation, the square root tensor ${\bm \gamma}'_{ab}$ is type $[1,1\cdots 1]$,
and then we can get the equations of motion  for the tensor ${\bm \gamma}'$
\begin{equation}
G_{ab} + m^2 X_{ab}[{\bm \gamma}']=0\, .
\end{equation}
 Since the equations of motion are invariant under this translation, and considering
\begin{equation}
X_{ab}[{\bm \gamma}']=X_{ab}[{\bm \gamma}]-X_{ab}[\Delta]\, ,
\end{equation}
we get the equations of motion for ${\bm \gamma}$
\begin{equation}
G_{ab} + m^2X_{ab}[{\bm \gamma}]-m^2X_{ab}[\Delta]=0\, .
\end{equation}
Here $\Delta$ represents the translation, and it can be expressed as $\Delta_{ab}=\Phi \ell_a\ell_b$. Similarly, one gets the equations of motion
in the case of type $[31\cdots 1]$ with vanishing $\gamma_1$.
By this consideration, we get  equations of motion for all possible square root tensors.

\section{Energy conditions}\label{energy condi}
We have got  equations of motion for all possible square root tensors. However, the reference metric $f_{ab}$ and the effective energy momentum tensor $-X_{ab}$ might be not physically acceptable for these square root tensors. In this section
we study the properties of $X_{ab}$ in the null (or orthogonal) frames. For simplicity, we only discuss the problem in four dimensions. The case of higher dimensions is straightforward.

\subsection{Sygre type $[1,111]$}
When ${\bm \gamma}_{ab}$ is of Sygre type $[1,111]$, we have
\begin{equation}
f_{ab}= - 2(\alpha_1^2 + \alpha_2^2) \ell_{(a}n_{b)}+2\alpha_1 \alpha_2( \ell_a\ell_b + n_an_b) + \alpha_{3}^2x_a^{(1)}x_b^{(1)}+ \alpha_{4}^2x_a^{(2)}x_b^{(2)}\, .
\end{equation}
So $f_{ab}$ has the same Sygre type with ${\bm \gamma}_{ab}$. This is natural because they have diagonal forms in the orthogonal frame. It is also obvious that $f_{ab}$ (if non-degenerate) has Lorentzian signature automatically.

After some calculation, we find
\begin{equation}
\label{condition1}
-X_{ab}\ell^a\ell^b = -X_{ab}n^an^b = -\alpha_2\big[c_1 + 2 c_2 (\alpha_3 + \alpha_4) + 6 c_3\alpha_3  \alpha_4\big]\, .
\end{equation}
Since we have chosen that $\ell^a$ and $n^a$ are both future pointing, the right hand side of the above equation should be non-negative if the null energy condition has to be satisfied. This gives a condition on the coefficients $c_i$ and
the eigenvalues of the principal directions in the theory. If we choose $\alpha_i$s to be all positive, then $c_i$s have to be nonpositive.

\subsection{Sygre type $[21 1]$}

In the case where ${\bm \gamma}_{ab}$ is of type $[211]$, it is not hard to find the reference metric to be
\begin{equation}
f_{ab}= - 2\beta_1^2 \ell_{(a}n_{b)}+2\beta_1 \beta_2  \ell_a\ell_b+ \beta_{3}^2x_a^{(1)}x_b^{(1)}+ \beta_{4}^2x_a^{(2)}x_b^{(2)}\, .
\end{equation}
Obviously, if $\beta_1=0$, the reference metric is no longer of type $[211]$. This point has been mentioned in the previous section. If $\beta_1\ne 0$, we can rescale the null frame such that
$f_{ab}$ has standard form of type $[211]$. Calculation shows
\begin{equation}
-X_{ab}\ell^a\ell^b= 0\, ,
\end{equation}
and
\begin{equation}
 -X_{ab}n^an^b = -\beta_2\big[c_1 + 2 c_2 (\beta_3 + \beta_4) + 6 c_3\beta_3  \beta_4\big]\, .
\end{equation}
So $\beta_2$ can not be $-1$, if the $c_i$s,  $\beta_3$, and $\beta_4$ have been chosen to satisfy the condition in (\ref{condition1}).

\subsection{Sygre type $[31]$}
If the square root tensor is of type $[31]$, we have
\begin{equation}
f_{ab}= - 2\gamma_1^2 \ell_{(a}n_{b)}+  \ell_a\ell_b+ 4\gamma_1 \ell_{(a}x_{b)}+ \gamma_{2}^2x_a^{(2)}x_b^{(2)}\, .
\end{equation}
The reference metric (if  non-degenerate) also has Lorentzian signature.  In the case $\gamma_1=0$, $f_{ab}$ has
Sygre type $[21 1]$ and has well defined pseudoinverse. In the case where $\gamma_1$ is nonvanishing, we can perform a Lorentz transformation of the null frame
and write $f_{ab}$ in the standard form of Sygre type $[211]$. In a word, the reference metric is always of type $[211]$ with $\beta_3=0$.

The effective energy momentum tensor satisfies
\begin{equation}
-X_{ab}\ell^a\ell^b=0\, ,
\end{equation}
and
\begin{equation}
-X_{ab}n^an^b = 2\big[ c_2 + 3c_3\gamma_3\big]\, .
\end{equation}
So the null energy condition will be violated if $c_2<0$, $c_3<0$, and $\gamma_3\ge 0$. This suggests that type $[31]$ square root tensor is not physically preferred in classical theory.

\subsection{Sygre type $[z\bar{z}11]$}
When the square root tensor has complex eigenvalues, or is of type $[z\bar{z}11]$, the reference metric has the following form
\begin{equation}
f_{ab}=-2 (\sigma_1^2-\sigma_2^2)\ell_{(a}n_{b)} + 2\sigma_1\sigma_2( \ell_a\ell_b -n_an_b) + \sigma_3^2 x^{(1)}_ax^{(1)}_b+ \sigma_4^2 x^{(2)}_ax^{(2)}_b\, .
\end{equation}
Generally,  $f_{ab}$ has Lorentzian signature. However,  after some calculation, we find that the energy momentum tensor satisfies that
\begin{equation}
-X_{ab}\ell^a\ell^b = \sigma_2\big[c_1 + 2c_2(\sigma_3+\sigma_4) + 6 c_3 \sigma_3\sigma_4 \big]\, ,
\end{equation}
and
\begin{equation}
-X_{ab}n^an^b = -\sigma_2\big[c_1 + 2c_2(\sigma_3+\sigma_4) + 6 c_3 \sigma_3\sigma_4 \big]\, .
\end{equation}
This means the null energy condition is violated by this type of square root tensors. In other words, when the square root tensor has
complex eigenvalues, the null energy condition can not be fulfilled in general.

\section{The Birkhoff type theorem of the solutions with a rank-$n$ degenerate reference metric}\label{Birk theo}
In this section,
we first discuss the square root tensor  of a  rank-$n$ degenerate reference metric and prove a Proposition,
and then the Birkhoff type theorem of the solutions of equations of motion with a rank-$n$ degenerate reference metric.

\subsection{The square root tensor  of a rank-$n$ degenerate reference metric}\label{Prop}

Assume that the $d-$dimensional spacetime $(M, g_{ab})$ can be foliated by a family of $n$-dimensional spacelike surfaces ($d=n+2$), and the metric can be expressed as
\begin{equation}
\label{metricdecomp}
g_{ab}=-\ell_an_b -n_a\ell_b +\sigma_{a b}\, ,
\end{equation}
where $\ell_a$ and $n_a$ are two null vector fields which are normal to the $n$-dimensional spacelike surfaces. This means
\begin{equation}
\ell_a\ell^a=n_an^a=\ell^a\sigma_{ab}=n^a\sigma_{ab}=0\, .
\end{equation}
Further, we assume that
\begin{equation}
\ell_an^a=-1\, .
\end{equation}
Here, the ``reference metric" $f_{ab}$ is assumed to satisfy that
\begin{equation}
\label{condition}
f_{ab}=\sigma_{a}{}^c\sigma_b{}^df_{cd}\, .
\end{equation}

\noindent{\bf Proposition} {\it `` For the spacetime in which the metric has a decomposition (\ref{metricdecomp}), and the reference metric $f_{ab}$ is a non-degenerate tensor on the spacelike $n$-surface, the symmetric tensor ${\bm \gamma}_{ab}$ can only has the following forms
\begin{equation}
\label{P1}
{\bm \gamma}_{ab}= x \ell_a\ell_b+ \bar{{\bm \gamma}}_{ab} \, ,\quad \mathrm{or} \qquad {\bm \gamma}_{ab}= y n_an_b+ \bar{{\bm \gamma}}_{ab}\, ,
\end{equation}
where $x$ and $y$ are  arbitrary  functions on the spacetime and $\bar{{\bm \gamma}}_{ab}=\sigma_a{}^c\sigma_b{}^d{\bm \gamma}_{cd}$ ."
}

The proof of this proposition can be found in the  appendix \ref{Appendix}.  It is easy to find that the rank of ${\bm \gamma}$ is larger than $f$. According to the discussion in previous sections, we know that generally the Moore-Penrose inverse of this square root tensor does not exist unless $x=0$ and $y=0$.

The square root tensors in (\ref{P1})  are obviously degenerate along $\ell^a$ or $n^a$ directions. By considering the symmetry of the theory as in section 3.3, we can get the equations of motion for this ${\bm \gamma}_{ab}$. The final equations
are just the equations (\ref{eom1})   with ${\bm \gamma}_{ab}$ replaced by the tensor $\bar{{\bm \gamma}}_{ab}$ in eq.(\ref{P1}). In a word,  although the square root tensors have general forms in (\ref{P1}),  the dynamical parts of them are $\bar{{\bm \gamma}}_{ab}$, and the null parts of the tensor are gauge freedoms which can not be determined by the equations of motion. Of course, the null parts of the square root tensor can not effect the physics. This will be emphasized in the following Birkhoff-type theorem.

\subsection{The Birkhoff type theorem }
In this subsection, we will assume that the $d$-dimension spacetime $(M,g_{ab})$ has a symmetry of a maximally symmetric space with dimension $n=d-2$, and the metric $g_{ab}$ can be expressed as
\begin{equation}
\label{maxsymm}
g_{ab}=h_{ab}+ \sigma_{ab}\, , 
\end{equation}
where
\begin{equation}
h_{ab}=h_{AB}(y)(\mathrm{d}y^A)_a(\mathrm{d}y^B)_b
\end{equation}
can be viewed as the metric on a two dimensional Lorentz manifold $(M_2, h_{ab})$, and
\begin{equation}
\sigma_{ab}= r^2(y)q_{ij}(\mathrm{d}z^i)_a(\mathrm{d}z^j)_b\, .
\end{equation}
In the above equation, $q_{ij}\mathrm{d}z^i\mathrm{d}z^j$ is the standard metric of the $n-$dimensional  maximally symmetric space with sectional curvature $K$. The ``reference metric" $f_{ab}$ can be selected as
\begin{equation}
f_{ab}=r_0^2 q_{ij}(\mathrm{d}z^i)_a(\mathrm{d}z^j)_b\, ,
\end{equation}
where $r_0$ is a constant. By this selection, we can get the square root tensor
\begin{equation}
{\bm \gamma}_{ab}= \bar{{\bm \gamma}}_{ab}  + \mathrm{null~parts}\, ,
\end{equation}
where
\begin{equation}
\bar{{\bm \gamma}}_{ab}=r r_0q_{ij} (\mathrm{d}z^i)_a(\mathrm{d}z^j)_b\, .
\end{equation}
As we have discussed before, only $\bar{{\bm \gamma}}_{ab}$ is dynamical, and the null part does not appear in the equations of motion.
Given this, it is easy to find that
\begin{equation}
({\bm \gamma}^k)_a{}^b=\Big(\frac{r_0}{r}\Big)^kq_{i}{}^j(\mathrm{d}z^i)_a\Big(\frac{\partial}{\partial z^j}\Big)^b\, ,
\end{equation}
and
\begin{eqnarray}
&&\mathcal{U}_1= \frac{n r_0}{r}\, ,\nonumber\\
&&\mathcal{U}_2=\frac{n(n-1)r_0^2}{r^2}\, ,\nonumber\\
&&\mathcal{U}_3=\frac{n(n-1)(n-2)r_0^3}{r^3}\, ,\nonumber\\
&&\mathcal{U}_4=\frac{n(n-1)(n-2)(n-3)r_0^4}{r^4}\, \,\nonumber\\
&&\cdots\cdots
\end{eqnarray}
So the only non-vanishing $\mathcal{U}_k$s are cases with $k\le n$, and we have
\begin{equation}
\mathcal{U}_k=\frac{n!c_k}{(n-k)!}\Big(\frac{r_0}{r}\Big)^{k}\, ,\qquad k=0\, ,\cdots\, ,n\, .
\end{equation}
The nontrivial components of the tensor $X_{ab}$ in eq.(\ref{Xab}) can be put into a form
\begin{eqnarray}
-2X_{AB}&=&\Big[\sum_{k=0}^{n}\frac{n!c_k}{(n-k)!}\Big(\frac{r_0}{r}\Big)^{k}\Big]h_{AB}\, ,
\end{eqnarray}
and
\begin{eqnarray}
-2 X_{ij}&=&\Big[ \sum_{k=0}^{n}\frac{(n-1)!c_k}{(n-1-k)!}\Big(\frac{r_0}{r}\Big)^{k}\Big]r^2q_{ij}\, .
\end{eqnarray}
For the spacetime (\ref{maxsymm}), the nontrivial components of Einstein tensor is simply
\begin{eqnarray}
G_{AB}=-n\frac{D_ArD_Br}{r} - \Big[\frac{1}{2}n(n-1)\frac{K-D_CrD^Cr}{r^2}-n\frac{D_CD^Cr}{r}\Big]h_{AB}\, ,\nonumber\\
G_{ij}=\Big[-\frac{1}{2} {}^{(2)}\!R -\frac{1}{2}(n-1)(n-2)\frac{K-D_CrD^Cr}{r^2} + (n-1)\frac{D_CD^Cr}{r}\Big]r^2q_{ij}\, ,\nonumber\\
\end{eqnarray}
and the scalar curvature of the spacetime is given by
\begin{equation}
R={}^{(2)}\!R -2n\frac{D_CD^Cr}{r} +n(n-1)\frac{K-D_CrD^Cr}{r^2}\, ,
\end{equation}
where ${}^{(2)}\!R$ is the scalar curvature of the two dimensional Lorentz manifold $(M_2, h_{ab})$, and $D_A$ is the  covariant derivative associated with the metric $h_{ab}$ along the natural basis of $y^A$.
Based on the above equations, we find the equations of motion reducing to
\begin{eqnarray}
\label{ABcomp}
&&-n\frac{D_AD_Br}{r} - \Big[\frac{1}{2}n(n-1)\frac{K-D_CrD^Cr}{r^2}-n\frac{D_CD^Cr}{r}\Big]h_{AB}\nonumber\\
&&-\frac{1}{2}m^2\Bigg[\sum_{k=0}^{n}\frac{n!c_k}{(n-k)!}\Big(\frac{r_0}{r}\Big)^{k}\Bigg]h_{AB}=0\, ,
\end{eqnarray}
and
\begin{eqnarray}
\label{ijcomp}
&&- {}^{(2)}\!R -(n-1)(n-2)\frac{K-D_CrD^Cr}{r^2} + 2(n-1)\frac{D_CD^Cr}{r}\nonumber\\
&&-m^2\Bigg[ \sum_{k=0}^{n}\frac{(n-1)!c_k}{(n-1-k)!}\Big(\frac{r_0}{r}\Big)^{k}\Bigg]=0\, . 
\end{eqnarray}
Contracting Eq.(\ref{ABcomp}) with $D^{B}r$ and subtracting the trace part of Eq.(\ref{ABcomp})  multiplied by $D_Ar$ (here, we have assumed that $r$ is not a constant function), we find
\begin{eqnarray}
&&D_{A}\Bigg\{n r^{n-1}\big(K- D_CrD^Cr\big) + m^2 r^{n+1}\Bigg[ \sum_{k=0}^{n}\frac{n!c_k}{(n+1-k)!}\Big(\frac{r_0}{r}\Big)^{k}\Bigg]\Bigg\}=0\, . \nonumber \\
\end{eqnarray}
This result tells us
\begin{eqnarray}
\label{KDRSquare}
&& m^2  \Bigg[ \sum_{k=0}^{n}\frac{(n-1)!c_k}{(n+1-k)!}\Big(\frac{r_0}{r}\Big)^{k}\Bigg]+\frac{ \big(K- D_CrD^Cr\big)}{r^2} =\frac{M}{r^{n+1}}\, ,
\end{eqnarray}
where $M$ is a constant. Substituting this relation into Eq.(\ref{ijcomp}) and the trace part of Eq.(\ref{ABcomp}), we have
\begin{equation}
\frac{D_CD^Cr}{r}=\frac{(n-1)M}{r^{n+1}} -m^2\Bigg[ \sum_{k=0}^{n}\frac{(n-1)!(k-2)c_k}{(n+1-k)!}\Big(\frac{r_0}{r}\Big)^{k}\Bigg] \, ,
\end{equation}and
\begin{equation}
{}^{(2)}\!R=\frac{n(n-1)M}{r^{n+1}} - m^2\Bigg[ \sum_{k=0}^{n}\frac{(n-1)!(k-1)(k-2)c_k}{(n+1-k)!}\Big(\frac{r_0}{r}\Big)^{k}\Bigg] \, .
\end{equation}
By using the above relations, we find that the scalar curvature of the spacetime can be expressed as
\begin{equation}
\label{curvaturespacetime}
R= - m^2 \sum_{k=0}^{n}\frac{(n-1)!(n-k+2)}{(n-k)!}c_k\Big(\frac{r_0}{r}\Big)^{k}\, .
\end{equation}
Thus,the scalar curvature of the spacetime is divergent at the points of $r=0$.
Again, with these relations,  Eq.(\ref{ABcomp}) tells us an important relation\begin{equation}
\label{importantrelation}
D_AD_Br -\frac{1}{2}(D_CD^Cr)h_{AB}=0\, .
\end{equation}
In the spacetime $(M, g_{ab})$, let  us define a vector field $\xi^a$ by
\begin{equation}
\xi_a=\xi_A (\mathrm{d}y^A)_a \, ,
\end{equation}
where
\begin{equation}
\xi_A=\epsilon_{AB}D^Br\, ,
\end{equation}
and $\epsilon_{AB}$ are the components of the Levi-Civita tensor in $(M_2, h_{ab})$. By using Eq.(\ref{importantrelation}), it is not hard to find
\begin{equation}
D_A\xi_B+D_B\xi_A=0\, ,
\end{equation}
which lead to
\begin{equation}
\nabla_a\xi_b + \nabla_b\xi_a=0\, .
\end{equation}
So $\xi^a$ is a Killing vector of the spacetime. Furthermore, we have that
\begin{equation}
\xi_a\xi^a=\xi_A\xi^A= \epsilon_{AB}\epsilon^{AC}D^Br D_Cr=-D_ArD^Ar\, .
\end{equation}
We can also define a vector field $\chi^a$ by
\begin{equation}
\chi_a= D_Ar (\mathrm{d} y^A)_a=(\mathrm{d}r)_a\, ,
\end{equation}
and then we have $\chi_a\xi^a=0$,  as well as,
\begin{equation}
\chi_a\chi^a=D_ArD^Ar\, .
\end{equation}
In the region where $D_ArD^Ar>0$, the vector field $\chi^a$ is spacelike and $\xi^a$ is timelike. So, in this region, the spacetime is stationary.

Assume that $t$ is the parameter of the Killing orbit, then we can express the Killing vector as
$\xi^a=(\partial/\partial t)^a$. In the region $D_ArD^Ar>0$, the function $r$ can be viewed as a coordinate, and the dual vector of
this coordinate basis, i.e, $(\partial/\partial r)^a$, is just $\chi_a=(\mathrm{d}r)_a$.  So we can construct a coordinate system $\{t,r,z^i\}$. In this coordinate system, the nontrivial components of the metric $g_{ab}$ can be expressed as
\begin{eqnarray}
&&g_{tt}=g_{ab}\Big(\frac{\partial}{\partial t}\Big)^a\Big(\frac{\partial}{\partial t}\Big)^b=-D_ArD^Ar\, ,\nonumber\\
&&(g_{rr})^{-1}=g^{rr}=g^{ab}(\mathrm{d}r)_a(\mathrm{d}r)_b=D_Ar D^Ar\, ,\nonumber\\
&&g_{ij}=r^2q_{ij}\, .
\end{eqnarray}
Substituting Eq.(\ref{KDRSquare}), we find
\begin{eqnarray}
-g_{tt}&=&K- \frac{M}{r^{n-1}} + m^2r^2\Bigg[\sum_{i=0}^{n}\frac{(n-1)!c_i}{(n+1-i)!} \Big(\frac{r_0}{r}\Big)^i\Bigg]\, . 
\end{eqnarray}
So without matter field, the solution (\ref{maxsymm}) is  Schwarzschild-type if the function $r$ is not a constant. The above deduction is the typical procedure to prove
the so-called Birkhoff theorem.

If we consider the cases where only $c_0\, ,c_1\, ,\cdots\, ,c_4$ are relevant, we get
\begin{eqnarray}
-g_{tt}&=&K- \frac{M}{r^{n-1}} + m^2\Big[\frac{c_0}{n(n+1)} r^2+\frac{ c_1}{n}r_0 r  +  c_2  r_0^2\nonumber\\
&&+ c_3 \frac{ (n-1)r_0^3}{r}
+ c_4  \frac{(n-1)(n-2)r_0^4}{r^2}\Big]\, .
\end{eqnarray}
This is just the static solutions found  in references~\cite{Davison:2013jba, Blake:2013bqa, Cai:2014znn}. Our derivation does not include any matter in the action. Other cases including certain type of matter, for example, an electromagnetic field, can be considered and it is trivial to obtain similar Birkhoff   theorems. It is similar to the generalized Birkhoff theorem for Einstein gravity
in electromagnetic vacuum (for example, see~\cite{Kodama:2003kk}).

\subsection{Solutions with constant radius function}
In the above discussion, we focused on the cases where $r$ is not a constant function. To complete the discussion, let us study the possible solutions with constant radius function. If $r=r_c$ is a constant, from Eqs.(\ref{ABcomp}) and (\ref{ijcomp}), it is easy to find that
\begin{equation}
{}^{(2)}\!R=m^2\Bigg[\sum_{k=0}^{n}\frac{(n-1)!(k-2)}{(n-k)!}c_k\Big(\frac{r_0}{r_c}\Big)^{k}\Bigg]\, .
\end{equation}
This equation implies that $(M_2, h_{ab})$ must be a constant curvature space. So the spacetime $(M, g_{ab})$ is a direct product of two constant curvature spaces if the solutions with constant $r$ exist. At the same time, we have that
\begin{equation}
\frac{K}{r_c^2}=-m^2\Bigg[\sum_{k=0}^{n}\frac{(n-2)!}{(n-k)!}c_k\Big(\frac{r_0}{r_c}\Big)^{k}\Bigg]\, .
\end{equation}
So it is possible to find some nontrivial solutions with constant $r$. In the case $n=2$ (or $d=4$), we have
\begin{itemize}
\item[(1).] If $c_0=0$, we have
\begin{equation}
r_c= - \frac{K+  c_2 m^2 r_0^2}{2c_1 m^2 r_0}\, .
\end{equation}
This suggests that $K$ must be positive when $c_1<0$ and $c_2<0$.

\item[(2).] If $c_0\ne 0$, we have positive $r_c$ when (we also assume that $c_1<0$ and $c_2<0$)
\begin{equation}
\frac{2 c_0K}{m^2 r_0^2}\le c_1^2- 2 c_0 c_2\, .
\end{equation}
In the case $c_0\ge 0$, we can chose nonpositive $K$. While, in the case $c_0<0$ and $c_1^2\ge 2 c_0c_2$, we can chose $K$ to be nonnegative.
\end{itemize}
By selecting coordinates $\{t,x\}$ in the two dimensional space $(M_2, h_{ab})$, we can get a solution
\begin{equation}
\label{newsolution}
g=-f(x)\mathrm{d}t^2 + f^{-1}(x)\mathrm{d}x^2 + r_c^2 q_{ij}\mathrm{d}z^i\mathrm{d}z^j\, .
\end{equation}
The function $f(x)$ is given by
\begin{equation}
f(x)= a + b x +cx^2\, ,
\end{equation}
where $a$ and $b$ are two integral constants, and
\begin{equation}
c=-\frac{1}{2}m^2\Bigg[\sum_{k=0}^{n}\frac{(n-1)!(k-2)}{(n-k)!}c_k\Big(\frac{r_0}{r_c}\Big)^{k}\Bigg]\, .
\end{equation}
This is a new solution of this theory. Of course, one can get more solutions by choosing different form of $h_{AB}$ in the two dimensional space $(M_2, h_{ab})$.
In fact, they are Nariai-Bertotti-Robinson-type solutions. Usually Nariai-Bertotti-Robinson-type solution is interesting when cosmological constant or electromagnetic field were presented. Here, we have constructed such kind of solution by general massive terms of gravity. If electromagnetic field is considered, one can get more general Nariai-Bertotti-Robinson-type solutions.

The scalar curvature of the spacetime now has a form
\begin{equation}
 R= - m^2 \sum_{k=0}^{n}\frac{(n-1)!(n-k+2)}{(n-k)!}c_k\Big(\frac{r_0}{r_c}\Big)^{k}\, .
\end{equation}
Which is just the expression (\ref{curvaturespacetime}) except that  $r$ is replaced by $r_c$.
This result means that the scalar curvature of the spacetime is a constant. It is also easy to find that there is  no singularity  in this spacetime.

\section{Conclusion and Discussion}

For the dRGT massive gravity theory, the equations of motion in the case with a degenerate reference metric are derived. While in cases with a non-degenerate reference metric the variation of the action includes introducing the inverse of the square root of the reference metric, i.e. the square root tensor, the procedure in degenerate cases is done by applying the generalized Moore-Penrose pseudoinverse of the square root tensor. However, when the generalized Moore-Penrose pseudoinverse does not exist, one need first to use the uncertainty, which we found to be a translation, of the square root tensor, to convert it to a form which has a Moore-Penrose pseudoinverse, then carry out the variation of the action. We also found that for some types of the square root tensor the null energy condition of the effective energy momentum tensor will be violated and therefore these types are physically unacceptable. In a word, the equations of motion for an arbitrary reference metric have been found. The equations of motion can be applied to many problems, for example the stability problems, 
which have been discussed in the non-degenerate cases by the authors of  \cite{Kodama:2013rea,Arraut:2015eta,Arraut:2015dva}.

Additionally, we prove a Birkhoff type theorem for a certain type of reference metric (thus for a certain type of the square root tensor according to the proposition we came up with) in some spacetime with a maximally symmetric subspace. The theorem rules out any solutions other than those found in the paper \cite{Cai:2014znn}. In order to find more general solutions one should consider other types of reference metrics, or assume some different structure of the spacetime.

It should be pointed out that the theorem should be understood as the uniqueness of the metric $g_{ab}$ under the condition of maximal symmetry (and the condition that radius function $r$ is not a constant).  However,  the square root tensor ${\bm \gamma_{ab}}$
can not be determined by this condition. It can include an arbitrary null part. The uniqueness of the metric is highly dependent on our analysis of the equations of motion in the case where the reference metric is degenerate.
If the equations of motion with  non-degenerate reference metric were naively used for the degenerate case,  one may find some solutions which break the Birkhoff theorem.  In the cases where the  generalized Moore-Penrose
inverses of the square root tensor do not exist, the massive gravity has such   a gauge freedom.  It is   interesting and deserving to study in future and find whether there is a deeper reason hidden behind this fact.

Another point which should be pointed out is that the solution (\ref{maxsymm}) always has a singularity even in the cases with a vanishing (black hole) mass parameter. This can be found from the scalar curvature of the spacetime, i.e., Eq.(\ref{curvaturespacetime}). Calculation shows that the  Kretschmann scalar of the spacetime is always divergent when the function $r$ approaches zero. So it seems that these solutions have no regular vacuums. This is very different from the solutions in Einstein gravity.
However, the solutions (\ref{newsolution}) with constant $r$ is regular. The physics on these Nariai-Bertotti-Robinson-type spacetimes have not been investigated up to date, and  it  needs  to be further studied.

\section*{Acknowledgement}
This work was supported in part by the National Natural Science Foundation of China with grants
No.11205148 and No.11235010. This work was also Supported by the Open Project Program of State Key Laboratory of Theoretical Physics, Institute of Theoretical Physics,
Chinese Academy of Sciences, China (No.Y5KF161CJ1). YLZ thanks support from MOST grant(No. 104-2811- M-002-080). LMC would like to thank  Rong-Gen Cai,  Yapeng Hu, Hongsheng Zhang, and Zhiguang Xiao for their useful discussion.

 \appendix
\section{ The proof of  proposition on square root tensor}
\label{Appendix}

By the definition of the square root tensor,  it is easy to find that
$$V_a={\bm \gamma}_{ab}n^b\, ,\qquad W_a={\bm \gamma}_{ab}\ell^b$$
are two null vector fields on the spacetime $(M, g_{ab})$, i.e., $V_aV^a=0$ and $W_aW^a=0$, and satisfy
\begin{equation}
\label{vworth}
V_aW^a=0\, .
\end{equation}
Since two orthogonal null vectors must be proportional to each other, we have  that
\begin{equation}
W_a= c V_a\,,
\end{equation}
where $c$ is an arbitrary function on the spacetime $(M,g_{ab})$.

In the above, we have assumed that ${\bm \gamma}_{ab}$ is symmetric.
Generally, the symmetric tensor ${\bm \gamma}_{ab}$ can be decomposed as
\begin{eqnarray}
&&{\bm \gamma}_{ab}= \delta_{a}{}^c\delta_b{}^d {\bm \gamma}_{cd}\nonumber\\
&&=(-\ell_an^c-n_a\ell^c +\sigma_a{}^c)(-\ell_bn^d-n_b\ell^d +\sigma_b{}^d){\bm \gamma}_{cd}\nonumber\\
&&=\ell_a\ell_b{\bm \gamma}_{nn}+n_a n_b {\bm \gamma}_{\ell\ell}+(\ell_an_b+n_a \ell_b){\bm \gamma}_{n\ell}\nonumber\\
&&
-2\ell_{(a}\bar{V}_{b)}-2n_{(a} \bar{W}_{b)}
+\bar{{\bm \gamma}}_{ab}\, ,
\end{eqnarray}
where
\begin{eqnarray}
&&{\bm \gamma}_{\ell\ell}={\bm \gamma}_{ab}\ell^a\ell^b\, ,\qquad {\bm \gamma}_{\ell n}={\bm \gamma}_{ab}\ell^an^b\, ,\qquad
{\bm \gamma}_{nn}={\bm \gamma}_{ab}n^an^b\, ,\nonumber\\
&&\bar{V}_a=\sigma_a{}^b V_b\, ,\quad \qquad \bar{W}_a=\sigma_a{}^bW_b\, ,\quad\qquad  \bar{{\bm \gamma}}_{ab}=\sigma_a{}^c\sigma_b{}^d{\bm \gamma}_{cd}\, .
\end{eqnarray}
Based on this decomposition, it is not hard to find that
\begin{eqnarray}
V_a&=&{\bm \gamma}_{ab}n^b=-\ell_a {\bm \gamma}_{nn} - n_a{\bm \gamma}_{n\ell}+\bar{V}_a\, ,\nonumber\\
W_a&=&{\bm \gamma}_{ab}\ell^b=-n_a {\bm \gamma}_{\ell\ell} - \ell_a{\bm \gamma}_{n\ell}+\bar{W}_a\,,
\end{eqnarray}
so we have that
\begin{eqnarray}
&&V_aV^a=-2{\bm \gamma}_{nn}{\bm \gamma}_{n\ell}+\bar{V}_a\bar{V}^a=0\, ,\nonumber\\
&&W_aW^a=-2{\bm \gamma}_{\ell\ell}{\bm \gamma}_{n\ell}+\bar{W}_a\bar{W}^a=0\, ,\nonumber\\
&&V_aW^a=-{\bm \gamma}_{nn}{\bm \gamma}_{\ell\ell}-{\bm \gamma}_{n\ell}{\bm \gamma}_{n\ell}+\bar{V}_a\bar{W}^a=0\, .
\end{eqnarray}
On the other hand, we have  that
\begin{eqnarray}
&&{\bm \gamma}_a{}^c{\bm \gamma}_c{}^b
=-\ell_a\ell^b({\bm \gamma}_{nn}{\bm \gamma}_{n\ell}+{\bm \gamma}_{nn}{\bm \gamma}_{n\ell}-\bar{V}_c\bar{V}^c)\nonumber\\
&&-n_an^b({\bm \gamma}_{\ell\ell}{\bm \gamma}_{n\ell}+{\bm \gamma}_{\ell\ell}{\bm \gamma}_{n\ell}-\bar{W}_c\bar{W}^c)\nonumber\\
&&(-\ell_an^b-n_a\ell^b)({\bm \gamma}_{nn}{\bm \gamma}_{\ell\ell} +{\bm \gamma}_{n\ell}{\bm \gamma}_{n\ell}-\bar{V}_c\bar{W}^c)\nonumber\\
&&+(\ell_a\bar{W}^b+\bar{W}_a\ell^b){\bm \gamma}_{nn}+(n_a\bar{V}^b+\bar{V}_an^b){\bm \gamma}_{\ell\ell}\nonumber\\
&&+{\bm \gamma}_{n\ell}(\ell_a\bar{V}^b
+n_a\bar{W}^b+\bar{V}_a\ell^b +\bar{W}_an^b)\nonumber\\
&&-\ell_a \bar{V}^c\bar{{\bm \gamma}}_c{}^b-n_a\bar{W}^c\bar{{\bm \gamma}}_c{}^b-\bar{{\bm \gamma}}_a{}^c\bar{V}_c\ell^b -\bar{{\bm \gamma}}_a{}^c\bar{W}_cn^b\nonumber\\
&&-\bar{V}_a\bar{W}^b-\bar{W}_a\bar{V}^b + \bar{{\bm \gamma}}_a{}^c \bar{{\bm \gamma}}_c{}^b\, ,
\end{eqnarray}
so we get
\begin{eqnarray}
&&\ell^a{\bm \gamma}_a{}^c{\bm \gamma}_c{}^b
=\ell^b({\bm \gamma}_{nn}{\bm \gamma}_{\ell\ell} +{\bm \gamma}_{n\ell}{\bm \gamma}_{n\ell}-\bar{V}_c\bar{W}^c)\nonumber\\
&&+n^b({\bm \gamma}_{\ell\ell}{\bm \gamma}_{n\ell}+{\bm \gamma}_{\ell\ell}{\bm \gamma}_{n\ell}-\bar{W}_c\bar{W}^c)\nonumber\\
&&-{\bm \gamma}_{\ell\ell}\bar{V}^b-{\bm \gamma}_{n\ell}\bar{W}^b
+\bar{W}^c\bar{{\bm \gamma}}_c{}^b\, ,
\end{eqnarray}
and
\begin{eqnarray}
&&{\bm \gamma}_a{}^c{\bm \gamma}_c{}^bn_b
=n_a({\bm \gamma}_{nn}{\bm \gamma}_{\ell\ell} +{\bm \gamma}_{n\ell}{\bm \gamma}_{n\ell}-\bar{V}_c\bar{W}^c)\nonumber\\
&&+\ell_a({\bm \gamma}_{nn}{\bm \gamma}_{n\ell}+{\bm \gamma}_{nn}{\bm \gamma}_{n\ell}-\bar{V}_c\bar{V}^c)\nonumber\\
&&{\bm \gamma}_{nn}-\bar{W}_a-{\bm \gamma}_{n\ell}\bar{V}_a+\bar{{\bm \gamma}}_a{}^c\bar{V}_c\, .
\end{eqnarray}
Since we have assumed that $f_{ab}\equiv{\bm \gamma}_a{}^c{\bm \gamma}_{cb}$ satisfies condition (\ref{condition}), the above two equations imply the following algebraic equations
\begin{eqnarray}
{\bm \gamma}_{nn}{\bm \gamma}_{\ell\ell} +{\bm \gamma}_{n\ell}{\bm \gamma}_{n\ell}-\bar{V}_c\bar{W}^c=0\, ,\nonumber\\
{\bm \gamma}_{\ell\ell}{\bm \gamma}_{n\ell}+{\bm \gamma}_{\ell\ell}{\bm \gamma}_{n\ell}-\bar{W}_c\bar{W}^c=0\, ,\nonumber\\
-{\bm \gamma}_{\ell\ell}\bar{V}^b-{\bm \gamma}_{n\ell}\bar{W}^b+\bar{W}^c\bar{{\bm \gamma}}_c{}^b=0\, .
\end{eqnarray}
and
\begin{eqnarray}
{\bm \gamma}_{nn}{\bm \gamma}_{\ell\ell} +{\bm \gamma}_{n\ell}{\bm \gamma}_{n\ell}-\bar{V}_c\bar{W}^c=0\, ,\nonumber\\
{\bm \gamma}_{nn}{\bm \gamma}_{n\ell}+{\bm \gamma}_{nn}{\bm \gamma}_{n\ell}-\bar{V}_c\bar{V}^c=0\, ,\nonumber\\
-{\bm \gamma}_{nn}\bar{W}_a-{\bm \gamma}_{n\ell}\bar{V}_a+\bar{{\bm \gamma}}_a{}^c\bar{V}_c=0\, .
\end{eqnarray}
As we have mentioned before, $W^a$ is proportional to $V^a$, i.e., (\ref{vworth}), so the above equations reduce to
%
\begin{eqnarray}
&&xy +z^2-c\bar{V}_c\bar{V}^c=0\, ,\nonumber\\
&&2yz-c^2\bar{V}_c\bar{V}^c=0\, ,\nonumber\\
&&2xz-\bar{V}_c\bar{V}^c=0\, ,\nonumber\\
&&cx\bar{V}_a+z\bar{V}_a-\bar{{\bm \gamma}}_a{}^c\bar{V}_c=0\, ,\nonumber\\
&&y\bar{V}^b+cz\bar{V}^b-c\bar{V}^c\bar{{\bm \gamma}}_c{}^b=0\, ,
\end{eqnarray}
where
\begin{equation}
x={\bm \gamma}_{nn}\, ,\qquad y= {\bm \gamma}_{\ell\ell}\, ,\qquad z={\bm \gamma}_{n\ell}\, .
\end{equation}
The solutions to the above equations can be put into two classes which will be denoted by Case I and Case II.
\begin{itemize}

\item[(1).] Case I, where $\bar{V}_c=0$. We have the solutions
\begin{equation}
z=0\, ,\quad y=0\, , \quad x ~\mathrm{is} ~\mathrm{arbitrary}\, ,
\end{equation}
or
\begin{equation}
z=0\, ,\quad x=0\, , \quad y ~\mathrm{is} ~\mathrm{arbitrary}\, .
\end{equation}
These suggest that ${\bm \gamma}_{ab}$ has the expression
\begin{equation}
{\bm \gamma}_{ab}= x\ell_a\ell_b + \bar{{\bm \gamma}}_{ab}\, ,\qquad \mathrm{or}\qquad {\bm \gamma}_{ab}=yn_an_b + \bar{{\bm \gamma}}_{ab}\, .
\end{equation}
In this case,  we have that
\begin{equation}
{\bm \gamma}_a{}^c{\bm \gamma}_c{}^b=\bar{{\bm \gamma}}_a{}^c\bar{{\bm \gamma}}_c{}^b\, .
\end{equation}

\item[(2).] Case II, where $\bar{V}_c\ne 0$. Then $c\ne 0$, $x\ne 0$, $y\ne 0$, and $z\ne 0$,  we have
\begin{equation}
z=cx\, ,\qquad y=c^2 x\, ,\qquad \bar{V}_c\bar{V}^c =2 c x^2\, ,
\end{equation}
and
\begin{equation}
\bar{{\bm \gamma}}_a{}^c\bar{V}_c = 2c x \bar{V}_a\, .
\end{equation}
So $\bar{V}_a$ is an eigenvector of $\bar{{\bm \gamma}}_{a}{}^b$ with the eigenvalue $2c x$. Since $\bar{V}_a\bar{V}^a>0$, $c$ must be positive.
In this case, ${\bm \gamma}_{ab}$ can be expressed as
\begin{eqnarray}
&&{\bm \gamma}_{ab}=x(\ell_a+cn_a)(\ell_b + cn_b)\nonumber\\
&&- (\ell_a+cn_a)\bar{V}_b -(\ell_b+cn_b)\bar{V}_a +\bar{{\bm \gamma}}_{ab}\,.
\end{eqnarray}
By defining
\begin{equation}
u^a = \ell_a + cn_a\, ,\qquad v_a = \bar{V}_a\, ,
\end{equation}
we have
\begin{eqnarray}
{\bm \gamma}_{ab}=xu_au_b-2u_{(a}v_{b)}  +\bar{{\bm \gamma}}_{ab}\, ,
\end{eqnarray}
where
\begin{equation}
u_au^a=-2c\, ,\qquad v_av^a = 2 c x^2\,.
\end{equation}
This means that $u_a$ is a timelike vector field. ${\bm \gamma}_{ab}$ can be further put into another form
\begin{eqnarray}
{\bm \gamma}_{ab}=\lambda \hat{u}_a\hat{u}_b-2\lambda \hat{u}_{(a}\hat{v}_{b)}  +\bar{{\bm \gamma}}_{ab}\, ,
\end{eqnarray}
where
\begin{equation}
\hat{u}^a=u^a/\sqrt{2c} \, ,\qquad \hat{v}^a= v^a/\sqrt{2c x^2}\, ,\qquad \lambda =2c x\, ,
\end{equation}
and
\begin{equation}
\hat{u}_a\hat{u}^a=-1\, ,\qquad \hat{v}_a\hat{v}^a=1\, ,\qquad \hat{u}_a\hat{v}^a=0\, ,\qquad \bar{{\bm \gamma}}_a{}^c\hat{v}_c=\lambda \hat{v}_a\, .
\end{equation}
In this case, denoted by Case II,  we have
\begin{equation}
f_{ab}={\bm \gamma}_{a}{}^c{\bm \gamma}_{cb}=-\lambda^2\hat{v}_a\hat{v}_b+\bar{{\bm \gamma}}_{a}{}^c\bar{{\bm \gamma}}_{cb}\, .
\end{equation}
This equation implies
\begin{equation}
f_{ab}\hat{v}^a=0\, .
\end{equation}
However, since $\hat{v}^a$ is non-vanishing, $f_{ab}$ is a degenerate tensor on the $n-$dimensional spacelike surface. Of course, it can not be proportional to $\sigma_{ab}$. Because $f_{ab}$ is a tensor of the $n$-dimensional surface and degenerate along the direction of $\hat{v}^a$, in the case $n=2$,
it should has a form
\begin{equation}
f_{ab}=\alpha \hat{w}_a\hat{w}_b\, .
\end{equation}
where $w^a$ is tangent to the 2-dimensional surface and satisfies that
 \begin{equation}\hat{v}^a\hat{w}_a=0\, ,\qquad \hat{v}^a\hat{w}_a=1\, .
\end{equation}
\end{itemize}
Based on these discussions, we obtain the proposition in section \ref{Prop}.

\end{document}